\documentclass[%
reprint,
superscriptaddress,
nofootinbib,
amsmath,amssymb,
aps,
pre,
notitlepage,
showkeys]{revtex4-2}

\usepackage[colorlinks=true, urlcolor=blue, anchorcolor=blue, citecolor=blue,filecolor=blue,linkcolor=blue,menucolor=blue]{hyperref}

\usepackage{url}
\usepackage{graphicx}
\usepackage[export]{adjustbox}
\usepackage{mathtools}
\usepackage[T1]{fontenc}
\usepackage{amsmath}
\usepackage{amssymb}
\usepackage{stmaryrd}
\usepackage[utf8]{inputenc}
\usepackage{rotating}
\usepackage{comment}
\usepackage{url}
\usepackage{bbm}
\usepackage{soul}
\usepackage[normalem]{ulem}
\usepackage{tabularx,booktabs}
\usepackage{hyperref}  
\usepackage{subfigure}
\usepackage{xr}
\usepackage{tcolorbox}
\usepackage{paralist}
\usepackage{booktabs}
\usepackage{xcolor}
\usepackage{pifont}

\definecolor{myblue}{RGB}{119, 170, 218}
\definecolor{mygreen}{RGB}{44, 126, 51}

\newcommand{\la}{\langle}
\newcommand{\ra}{\rangle}

\begin{document}
\title{The Origins of Transient Bimodality} 

\author{Kaan \"Ocal$^{\dagger}$}
\affiliation{School of BioSciences, University of Melbourne, Parkville, Victoria 3052, Australia.}
\affiliation{School of Mathematics and Statistics, University of Melbourne, Parkville, Victoria 3052, Australia.}

\author{Augustinas Sukys$^{\dagger}$}
\affiliation{School of BioSciences, University of Melbourne, Parkville, Victoria 3052, Australia.}

\author{Aanjaneya Kumar}
\affiliation{The Santa Fe Institute, 1399 Hyde Park Road, Santa Fe, NM, 87501, USA.}

\author{James Holehouse}
\email{jamesholehouse1@gmail.com}
\affiliation{The Santa Fe Institute, 1399 Hyde Park Road, Santa Fe, NM, 87501, USA.}
\affiliation{School of Biology, Washington University in St. Louis, St. Louis, Missouri, USA.}

\begin{abstract}
    \noindent Many dynamical systems exhibit diverse modes of behavior. In biology, such modes can represent individual or cell fates. While the emergence of multimodality is commonly studied, transient bimodality is much less well understood. Under transient bimodality, a system moving from a well-defined initial to a final state transiently undergoes a bifurcation into multiple probability modes. This noise-driven phenomenon can significantly impact processes such as cell differentiation and speciation in the presence of changing environmental conditions. We detail a theoretical approach for understanding transient bimodality connecting results from ecology, optics, chemical reaction networks and cell biology, propose a ``minimal model'' of transient bimodality and derive a general criterion for its presence. We show that fast-to-slow dynamics can lead to transient bimodality in addition to the well-known case of slow-to-fast dynamics. Finally, we discuss the role of transient bimodality across the scientific literature, with emphasis on biochemical kinetics and gene regulation.
\end{abstract}

\maketitle
\def\thefootnote{$\dagger$}\footnotetext{Equal contribution.}\def\thefootnote{\arabic{footnote}}

\section*{Introduction}

\begin{figure*}[ht]
    \includegraphics[width=1\textwidth]{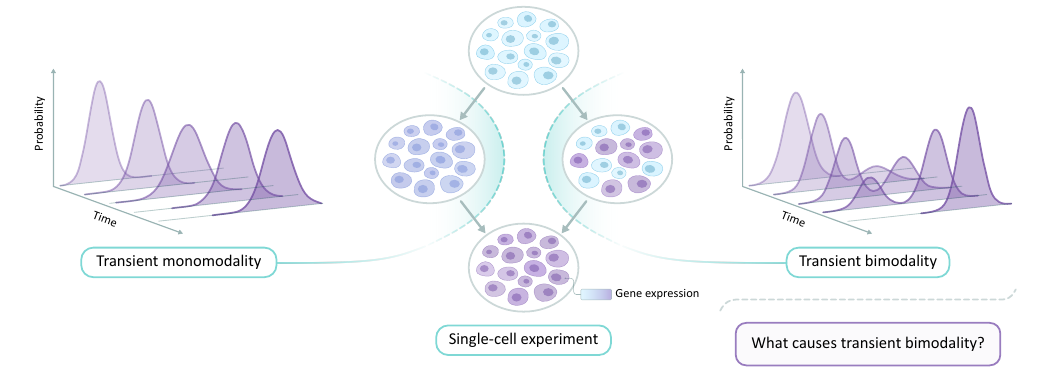}
    \caption{\textbf{The appearance of transiently bimodal behaviors.} In a single-cell perturbation experiment (middle panel), transient bimodality could manifest itself as an initial effective cell phenotype (characterized by a low expression level of an indicative gene; in blue) being replaced by a different phenotype (high expression state; in purple). A real example of transient bimodality similar to this can be observed in the expression of GAL genes in yeast cells upon the introduction of galactose, shown in ref.~\cite[Fig.~1]{stockwell2017living}. Rather than a smooth monomodal transition in expression level across the cell population (left panel), transient bimodality implies the occurrence of low- and high-expression subpopulations at intermediate times.}
    \label{fig:1}
\end{figure*}

\noindent Dynamical systems provide a unifying approach to study many scientific phenomena, from laser optics to ecology to cellular differentiation. One is often interested in the possible behaviors of a dynamical system, typically referred to as (stable) behavioral modes. Depending on how many such behavioral modes can be observed, deterministic systems are classified into monostable, bistable or multistable systems. In the presence of noise, behavioral modes are less clearly demarcated: they are typically understood as probability modes (see Fig.~\ref{fig:1}), basins of attraction in which individuals behave similarly. Noise can fundamentally alter the behavior of a dynamical system: multistable deterministic systems do not in general correspond to multimodal stochastic systems, or vice versa, a phenomenon that has been widely studied both theoretically \cite{qian2016framework,holehouse2026distinct,keizer2012statistical,vellela2007quasistationary,holehouse2020stochastic} and experimentally \cite{novick1957enzyme,blake2006phenotypic,wang2024transiently,levien2025slow,baptista2025bimodality,rattray2022bacterial}.

Since deterministic descriptions of dynamical systems poorly capture stochastic behavior \cite{fucho2025local,sardanyes2020noise,tomas2023semiclassical,hayashi2024deterministic,cobo2023emergence}, one often takes care to distinguish between stochastic and (mostly) deterministic systems. But what if a system is ``almost'' deterministic, exhibiting large fluctuations only over a transient period between two well-defined states? In this case, deterministic approaches can fail to predict system behavior, especially if the fluctuations occur in the vicinity of a bifurcation. This phenomenon, named transient bimodality, was observed in the 1980s and 1990s in the context of optical systems, chemical kinetics, and complex systems theory \cite{Baras1983,iwaniszewski1992transient,Iwaniszewski1994,iwaniszewski1995stochastic,broggi1984transient,broggi1985transient,dePasquale1985Apr,erdi2014stochastic,nowakowski2002thermal,agudov1999decay,lange1985study,van1987intrinsic,lemarchand1989experimental,doka2012stochastic}. In such systems, noise can  render a system highly susceptible to changing environmental conditions. Transient bimodality is difficult to predict from deterministic descriptions \cite{Baras1983,frankowicz1983transient}, frequently occurring in systems with dynamics that ``start slow and end fast'' \cite{nicolis1989exploring,frankowicz1983transient, frankowicz1984stochastic}. Given advances in real-time biological imaging \cite{fernandes2022synthetic} and sequencing \cite{chen2022live}, our study advances the understanding of transiently bimodal mechanisms which may become increasingly apparent in future empirical studies.

In biology, transient bimodality occurs when organisms or cells become transiently separated due to individual variability (see Fig.~\ref{fig:1}). At this point, environmental changes can lead to two subpopulations following very different trajectories, even if the population was previously evolving mostly deterministically. This can be observed in cell fate determination \cite{palani2012transient}, gene regulation \cite{maithreye2008delay} and antimicrobial responses in bacteria \cite{levien2025slow}; a well-studied example is found in the \textit{GAL} genes in budding yeast \cite{blake2003noise,blake2006phenotypic,venturelli2015population,stockwell2017living}, where protein distributions frequently exhibit transient bimodality upon galactose induction. This provides important information about the dynamics of transcription: bimodality arises due to the fact that some cells respond more rapidly than the rest of the population, which can significantly alter the fate of a cell.


Transient bimodality is also ubiquitous in ecology, where transient behavior underlies speciation events in population genetics and is commonly observed after environmental shocks \cite{abbott2021transients,vidiella2018exploiting,hastings2018transient,hastings2021effects,abbott2021mapping,vidiella2021habitat,koch2024biological}. Similar to the literature in cell biology \cite{levien2025slow}, environmental noise plays a crucial role around ecological transients \cite{abbott2021mapping,hastings2021effects}. Transient bimodality can act as a crossroads for a biological system, leading to rare, but impactful speciation events. Classical ecological systems are often treated as essentially deterministic with noise acting as a perturbation, which can neither explain nor model transient bimodality as we show in this paper.


Despite its role in shaping long-term biological behavior, transient bimodality has not often been studied in a biological context \cite{nicolis1989exploring}. A likely reason is that transient bimodality is a dynamical phenomenon and best observed in longitudinal studies, which can be difficult for both cell biology and ecology \cite{larson2011real,lammers2020matter,chen2022live,pomp2024transcription}. This highlights the need for a more thorough understanding of transient bimodality and the conditions in which it is likely to emerge. To this end, we present a minimal model of transient bimodality that allows us to study the basics of this phenomenon and derive a simple criterion based on first-passage times for when transient bimodality is likely to emerge. We illustrate our results using a power-law Markov death process, a scale-free model that can describe a variety of simple biological systems where transient bimodality has been observed. By studying the behavior of this model in the mono- and bimodal regimes, we uncover the core mechanisms that are responsible for transiently bimodal behavior.

\section*{Results}

\subsection{A Minimal Model of Transient Bimodality Suggests a Universal Criterion}
\label{sec:minimal}

\begin{figure*}[ht]
    \includegraphics[width=1\textwidth]{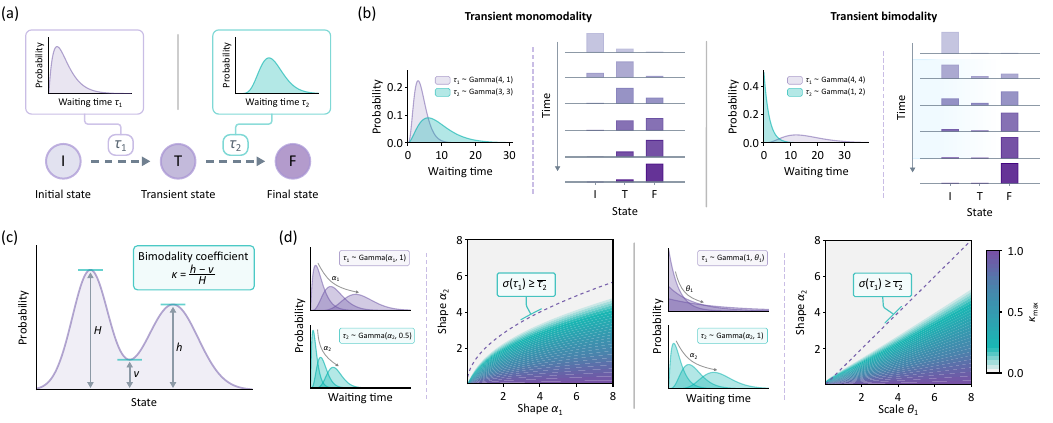}
    \caption{\textbf{Minimal model of transient bimodality.} (a) Schematic of the minimal model, where the random state-transition times $\tau_1$ ($I \to T$) and $\tau_2$ ($T \to F$) are described by two arbitrary waiting-time distributions. (b) Left: time evolution of the state distribution for $\tau_1 \sim \text{Gamma}(4, 1)$ and $\tau_2 \sim \text{Gamma}(3, 3)$; the system exhibits transiently monomodal dynamics. Right: time evolution of the state distribution for $\tau_1 \sim \text{Gamma}(4, 4)$ and $\tau_2 \sim \text{Gamma}(1, 2)$; the system exhibits transiently bimodal dynamics. The shaded region (in blue) highlights the time period over which transient bimodality is observed. (c) Schematic of the bimodality coefficient $\kappa$ that compares the heights of the two probability modes ($H$ and $h$) to the height of the valley between them ($v$). (d) Contour plots of the maximum bimodality coefficient, $\kappa_{\text{max}}$, encountered over the transient evolution of the system in $(\alpha_1, \alpha_2)$ and $(\theta_1, \alpha_2)$ spaces. In the left panel, $\alpha_1$ and $\alpha_2$ correspond to the shape parameters of the two gamma distributions used to characterize the waiting times $\tau_1$ and $\tau_2$, given by $\text{Gamma}(\alpha_1, 1)$ and $\text{Gamma}(\alpha_2, 0.5)$, respectively. On the right, the scale parameter $\theta_1$ of the first waiting time distribution is varied instead, so that $\tau_1$ and $\tau_2$ are given by $\text{Gamma}(1, \theta_1)$ and $\text{Gamma}(\alpha_2, 1)$. The light gray region indicates the absence of transient bimodality, and the purple dashed line delineates the region where the left-hand side is greater than the right-hand side in Eq.~\eqref{eq:crit}.}
    \label{fig:2}
\end{figure*}

\noindent We start by investigating a minimal model of transient bimodality, consisting of three states: an initial state $I$, a transient state $T$, and a final state $F$ (Fig.~\ref{fig:2}(a)). All individuals start in the initial state $I$. Each individual waits for a random amount of time $\tau_1 \sim p_1(\tau_1)$ before moving into the transient state $T$, and then waits for another random time $\tau_2 \sim p_2(\tau_2)$ before moving to the final state $F$. We assume that $\tau_1$ and $\tau_2$ are independent. The time-dependent probability of finding the system in any of the three states equals
\begin{align}
    p_I(t) &= P(\tau_1 > t), \\
    p_T(t) &= \int_0^t p_1(\tau_1) \, P(\tau_2 > t - \tau_1) \, d\tau_1, \\
    p_F(t) &= \int_0^t p_1(\tau_1) \, P(\tau_2 \leq t - \tau_1) \, d\tau_1,
\end{align}

\noindent where
\begin{align}
    P(\tau_i \leq s) &= \int_0^s p_i(\tau_i) \, d\tau_i, \\
    P(\tau_i > s) &= 1 - P(\tau_i \leq s)
\end{align}

\noindent are the cumulative distribution function of $\tau_i$ and its complement.

This system exhibits transient bimodality depending on the waiting-time distributions $p_1$ and $p_2$. Transient bimodality occurs if at some time $t^*$ there are more individuals in the initial and final states than in the transient state (Fig.~\ref{fig:2}(b)). For this to happen, we require some variability in the time $\tau_1$ spent in the initial state: more precisely, at time $t^*$, there must be enough individuals that have left the initial state (i.e., $\tau_1 < t^*$), and enough that have not (so $\tau_1 > t^*$). Furthermore, most of those individuals which have left the initial state must have moved on to the final state, i.e., $\tau_1 + \tau_2 < t^*$ for these individuals. Therefore, the \emph{spread} in dwelling times in the initial regime must be large compared to the \emph{typical} dwelling time in the transient regime. In mathematical terms, this suggests a criterion of the form
\begin{align}
    \sigma(\tau_1) &\gtrsim \overline{\tau_2}, \label{eq:crit}
\end{align}

\noindent where $\sigma(\tau_1)$ denotes the standard deviation of $\tau_1$, $\overline{\tau_2}$ the mean of $\tau_2$, and $\gtrsim$ inequality up to some constant factor. From this description, the time  $t^*$ of transient bimodality will be close to the median of $\tau_1$, when close to half of individuals can be found both in the initial state and in the transient or final states.

Eq.~\eqref{eq:crit} is not entirely quantitative: the standard deviation is but one measure of the spread, and the mean of $\tau_2$ is not always (close to) the typical value. However, we will see that Eq.~\eqref{eq:crit} is good at capturing transient bimodality in practice. Notably, our criterion is translation-invariant in $\tau_1$: adding a constant amount $\Delta t$ to the time each individual spends in $I$ only delays the time of bimodality by $\Delta t$. Thus, Eq.~\eqref{eq:crit} is agnostic to \emph{when} bimodality occurs. In contrast, increasing $\tau_2$ makes it less likely that we observe transient bimodality. Eq.~\eqref{eq:crit} is also scale-invariant in $\tau_1$ and $\tau_2$; as a result, it does not take into account how long bimodality is maintained for.

We now investigate how well Eq.~\eqref{eq:crit} predicts transient bimodality in our minimal model. To quantify the bimodality of a distribution, we use the bimodality coefficient $\kappa$ \cite{zhang2003bimodality,jia2020dynamical,holehouse2020stochastic} (see Fig.~\ref{fig:2}(c)). This geometric measure of bimodality compares the heights of the two probability modes to the height of the valley between them as
\begin{align}
    \kappa = \frac{h-v}{H}, \label{eq:kappa},
\end{align}
\noindent where $h$ is the height of the small mode, $v$ is the height of the valley between the modes, and $H$ is the height of the large mode. We define $\kappa = 0$ for distributions that have only a single mode. The bimodality coefficient has a minimal value of 0 (uni- or mono-
modality) and a maximal value of 1 (perfect bimodality). This makes $\kappa$ an effective \textit{order parameter}, distinguishing between the unimodal phase and the bimodal phase \cite{Haken1983Synergetics,goldenfeld2018lectures}. For our model, we compute the maximal bimodality coefficient 
\begin{align}
    \kappa_\mathrm{max} &= \sup_{t \geq 0} \kappa(t) \label{eq:kappa_max}
\end{align} 

\noindent to obtain a quantitative measure of bimodality in our system.

We numerically compare Eq.~\eqref{eq:crit} with Eq.~\eqref{eq:kappa_max} in the case where $\tau_1$ and $\tau_2$ are both Gamma distributed; this simple case allows us to capture a wide range of behaviors of our minimal model. As seen in Fig.~\ref{fig:2}(d), the region in which bimodality occurs (as measured by $\kappa$) is well demarcated by our criterion from Eq.~\eqref{eq:crit}. 
 

\subsection{Transient Bimodality in the Power-Law Death Process}
\label{sec:det}

\begin{figure*}[ht]
    \includegraphics[width=1\textwidth]{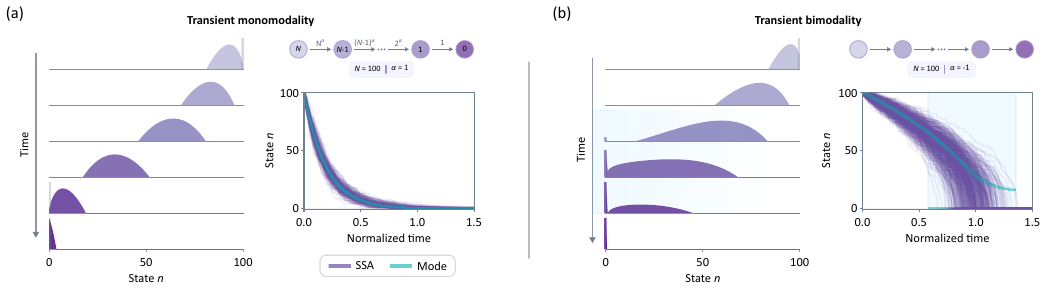}
    \caption{\textbf{Transient behavior of Markov death process with power-law rates.} (a) Transiently monomodal dynamics for $N=100$ and $\alpha=1$. Left: time evolution of the number distribution $P(n, t)$ (in log space), showing a single traveling mode of probability. The distributions $P(n, t)$ were computed by matrix exponentiation. Right: 100 sample stochastic trajectories of the Markov chain (purple) and the corresponding evolution of the single mode of $P(n, t)$ (turquoise). The horizontal axis represents time normalized by the mean first-passage time to reach the absorbing state. (b) Transiently bimodal dynamics for $N=100$ and $\alpha=-1$. Left: time evolution of the number distribution $P(n, t)$ (in log space), showing bimodal behavior at intermediate times. Right: 200 sample stochastic trajectories of the Markov chain (purple) and the corresponding evolution of the modes of $P(n, t)$ (turquoise). The shaded region (in blue) highlights the time period over which transient bimodality is observed. The gray vertical lines highlight the probability mass touching the boundaries of the state space, whereas the purple bars at $n=0$ emphasize the presence of a second mode at the absorbing state.} 
    \label{fig:3}
\end{figure*}

\noindent We illustrate the relevance of our results using the following Markov model:
\begin{align}\label{eq:chain}
    \boxed{N} \xrightarrow{N^\alpha} \boxed{N{-}1} \xrightarrow{(N-1)^\alpha}\cdots \boxed{2} \xrightarrow{2^\alpha} \boxed{1} \xrightarrow{1} \boxed{0}
\end{align}

\noindent This is a death process on $N{+}1$ states, where transition times from state $n$ to $n{-}1$ are independent and exponentially distributed. The transition rates follow a power law with an exponent $\alpha$, which renders our example scale-invariant, neglecting the inherent discreteness of this system. The chain has an absorbing state at $n=0$. Depending on $\alpha$, Eq.~\eqref{eq:chain} can exhibit three types of dynamics: (i) progressive slowing down ($\alpha>0$); (ii) constant speed ($\alpha = 0$); and (iii) progressive speeding up ($\alpha<0$). We exclude back-transitions (birth events) in \eqref{eq:chain} as these unnecessarily complicate the mathematical analysis of this model needed to get intuition for the conditions under which transient bimodality occurs \cite{holehouse2024first,hathcock2022asymptotic,smith2015general}.

Special cases of the Markov chain in Eq.~\eqref{eq:chain} are used to model various processes in physics and biology. For $\alpha = 0$, we obtain a Poisson point process starting at $N$ and ending at $0$, with first-passage times that are Erlang distributed, e.g.,~to model cell-cycle progression \cite{yates2017multi,perez2020effects}. When $\alpha = 1$, we obtain a linear death process, e.g.,~nuclear decay or mRNA/protein degradation in cells \cite{peccoud1995}. When $\alpha = 2$, we can model the number of polymers (shifted by $1$) in a simple polymerization process \cite{schnoerr2017approximation}, or molecule degradation in the dimerization process $2X \rightarrow X$ (with mass-action kinetics). Situations in which $0<\alpha<1$ correspond to decay with negative Hill function kinetics \cite{levitzki1969negative}. Finally, for $\alpha<0$ there are several useful examples: (i) degradation rates that increase as molecules are depleted \cite{zhao2016reversible}; (ii) inverse-density-dependent decline in ecology, in which the hazard of extinction increases as abundance decreases, a phenomenon widely associated with Allee effects \cite{courchamp1999inverse}; and (iii) chemical explosions \cite{Baras1983}.
While $n$ can be interpreted differently across these examples, in the following we interpret $n$ as the number of molecules/particles in a power-law death process.

The snapshot distribution of molecule numbers $P(n; t)$ in this system evolves according to the master equation
\begin{align}
    \frac{\mathrm{d}}{\mathrm{d}t} P(n, t) &= (n+1)^\alpha P(n+1, t) - n^\alpha P(n, t), \label{eq:cme}
\end{align}
where Eq.~\eqref{eq:cme} is defined for $1\leq n\leq N$, with $P(N+1,t)=0$, and the absorbing-state probability follows from normalization, $P(0,t)=1-\sum_{n=1}^{N}P(n,t)$.

Assuming the system starts in the state $N$ at $t = 0$, in Sec.~\ref{apdx:cme} we derive the following solution for \eqref{eq:cme}:
\begin{align}\label{eq:me_sol}
    P(n,t) = \frac{1}{n^\alpha }\sum_{i=n}^{N} i^\alpha e^{-i^\alpha t} \prod_{\substack{j=n\\ j\neq i}}^{N} \left(1-\frac{i^\alpha}{j^\alpha}\right)^{-1},\; \alpha,n \neq 0.
\end{align}

\noindent For the case of $\alpha=0$ degeneracy in the poles of the master equation in Eq.~\eqref{eq:cme} results in a modified form of $P(n,t)$, 
\begin{align}
    P(n,t) = \frac{t^{N-n}e^{-t}}{(N-n)!},\quad 1\leq n\leq N.
\end{align}
\noindent Since Eq.~\eqref{eq:me_sol} is rather opaque, we will later derive several simplifications to understand the behavior of this system. Fig.~\ref{fig:3} shows how $P(n, t)$ changes depending on $\alpha$. For $\alpha = 1$, the system remains unimodal at all times, while $\alpha = -1$ exhibits transient bimodality. Based on our discussion in the previous section, we can observe this behavior by looking at individual trajectories of the system (Fig.~\ref{fig:3}). Taking the transient regime to be around $n \approx 10$, for $\alpha = -1$ we see that the time $t_1$ it takes the system to reach this regime is highly variable. In contrast, once the system reaches $n = 10$, it quickly moves into the absorbing state $n = 0$, so that $t_2$ is comparatively small. This can be visualized by looking at the average velocity of the system at different stages (Fig.~\ref{fig:3}). The trajectories for $\alpha = 1$ show the precise opposite behavior, explaining the lack of bimodal behavior. 



\begin{figure*}[ht]
    \includegraphics[width=\textwidth]{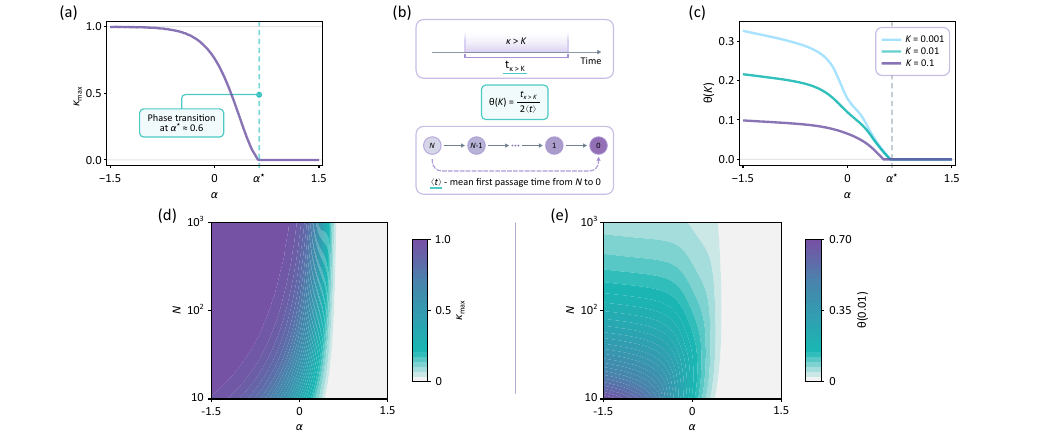}
    \caption{\textbf{Phase transition into transient bimodality.} (a) Plot of the maximum bimodality coefficient, $\kappa_{\text{max}}$, encountered over the transient evolution of the Markov chain, versus $\alpha$ for fixed $N=100$. The phase transition point is located near \smash{$\alpha^\star \approx 0.6$}, indicated by the dashed line. (b) Schematic of $\theta(K)$, a bimodality measure defined as the fraction of time a transient trajectory spends in a transiently bimodal regime (wherein $\kappa(t) > K$) between $t = 0$ and $t = 2\la t\ra$ (double of the mean first-passage time from the initial state to the final state). (c) Plot of $\theta(K)$ versus $\alpha$ for three different values of $K$. (d) Contour plot of the maximum bimodality coefficient, $\kappa_{\mathrm{max}}$, in $(\alpha,N)$ space. The light gray region indicates the absence of transient bimodality. (e) Contour plot of the fraction of time a trajectory is transiently bimodal, $\theta(K)$ (above the threshold $K=0.01$), in $(\alpha,N)$ space. The underlying number distributions $P(n, t)$ were computed using Krylov subspace methods implemented in \texttt{ExponentialUtilities.jl}.}
    \label{fig:4}
\end{figure*}

We next investigate for what values of $\alpha$ we observe transient bimodality. Fig.~\ref{fig:4}(a) shows the maximal bimodality coefficient $\kappa_\mathrm{max}$ as a function of $\alpha$. Treating this as an order parameter, we observe a second-order phase transition at a value of $\alpha^\star\approx 0.6$. We can compare this with another measure of bimodality, $\theta(K)$, which we define as the fraction of time for which $\kappa(t) > K$ between $t = 0$ and $t = 2\la t\ra$, twice the mean first-passage time to the final state (see Fig~\ref{fig:4}(b)). Fig.~\ref{fig:4}(c) shows that this quantity follows a qualitatively similar pattern to $\kappa_\mathrm{max}$ independently of $K$, with a transition around $\alpha \in [0.5,0.6]$ that weakly depends on the threshold value $K$. Lastly, Figs.~\ref{fig:4}(d) and (e) demonstrate that the properties of the transiently bimodal regime also depend on the initial state of the system $N$: as $N$ is increased, transient bimodality becomes more apparent and $\kappa_\mathrm{max}$ increases, but the relative amount of time over which it manifests tends to shrink and $\theta(K)$ decreases.

In an attempt to explain this phenomenon, we first use a deterministic approximation of our system \cite{Iwaniszewski1994,frankowicz1983transient,frankowicz1983transient} where we describe the state of the system as a continuous variable $x$, which corresponds to $n$ in the discrete formulation. In the deterministic limit, where the fluctuations around the mean value $\overline{x}(t)$ are small, the system is described by the rate equation 
\begin{align}\label{eq:mfmean}
    \frac{\mathrm{d}\overline{x}}{\mathrm{d}t} = -\overline{x}^\alpha = -\frac{\partial U(\overline{x})}{\partial \overline{x}},
\end{align}

\noindent with a deterministic potential 
\begin{align}
    U(\overline{x}) = \begin{cases}
        \frac 1 {1 + \alpha} \overline{x}^{1+\alpha}, &\alpha \neq -1,\\
        \log \overline{x}, &\alpha = -1.
    \end{cases}
\end{align}

The potential is increasing for all values of $\alpha$, with no local minima that could explain the second mode observed in transient bimodality (bistability would require two local minima corresponding to the two modes, see ref.~\cite{holehouse2026distinct}). The solution to \eqref{eq:mfmean} is given by
\begin{align}\label{eq:mean}
    \overline{x}(t) = 
    \begin{cases}
        (N^{1-\alpha}-(1-\alpha)t)^{1/(1-\alpha)}, &\alpha\neq 1,\\
        N \exp(-t), &\alpha = 1,
    \end{cases}
\end{align}

\noindent which does not qualitatively change during the switch to transient bimodality. 
This suggests that a deterministic description of our system cannot detect the occurrence of transient bimodality for $\alpha \lesssim 0.6$. In contrast, a simple stochastic treatment of this system explains this phenomenon to good accuracy. We explore this in the next section.

\subsection{Transient Bimodality can be Predicted from Snapshot Behavior}
\label{sec:snapshot}

\begin{figure*}[ht]
    \includegraphics[width=\textwidth]{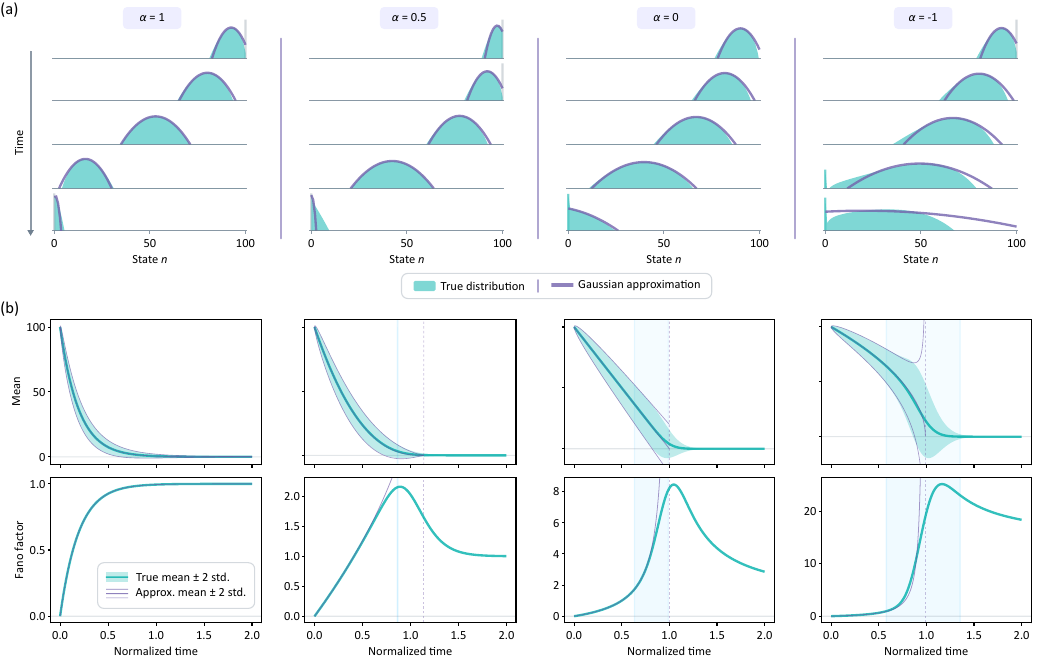}
    \caption{\textbf{Probing transient bimodality via mean-field theory and the moments of the number distribution.} (a) Testing the validity of the Gaussian approximation of the death process: each panel shows the time evolution of the number distribution (in log-space) for different values of $\alpha$, with $N=100$ in all cases. The shaded turquoise area corresponds to the true distribution, and the purple lines indicate the Gaussian approximation from Eq.~\eqref{eq:pxt}. The gray vertical lines highlight the probability mass touching the boundaries of the state space, whereas the turquoise bars at $n=0$ emphasize the presence of a second mode at the absorbing state. (b) Transient behavior of the system in terms of its mean and the Fano factor (for the same $N$ and $\alpha$ values). Top: time evolution of the mean, where the shaded band around the average trajectory covers $\pm$ 2 standard deviations. The purple lines (mean $\pm$ 2 standard deviations; Eqs.~\eqref{eq:mean} and \eqref{eq:sigma}) show the associated Gaussian approximation. The purple vertical dashed line indicates the point at which the mean-field theory breaks down for the given $\alpha$. The shaded region (in blue) highlights the time period over which transient bimodality is observed. Bottom: the corresponding time evolution of the Fano factor, $\mathrm{FF}(t)$. The horizontal axis represents time normalized by the mean first-passage time to reach the absorbing state.}
    \label{fig:5}
\end{figure*}

\noindent We next derive a Gaussian approximation of Eq.~\eqref{eq:cme} centered around the deterministic trajectory:
\begin{align}\label{eq:pxt}
    P(x,t) \approx \frac{1}{\sqrt{2\pi \sigma(t)^2}}\exp\left(-\frac{(x-\overline{x}(t))^2}{2\sigma(t)^2}\right),
\end{align}

\noindent where $\overline{x}(t)$ is given by Eq.~\eqref{eq:mean}, and $\sigma(t)$ is the standard deviation, which we have to compute. Following ref.~\cite{Baras1983}, we approximate $\sigma(t)$ using the WKB method (see Sec.~\ref{sec:mft}) to obtain
\begin{align} \label{eq:sigma}
    \sigma(t)^2 = 
    \begin{cases}
    \frac{1}{1-2\alpha}\left[N \left( \frac{\overline{x}(t)}{N} \right)^{2\alpha} - \overline{x}(t) \right], &\alpha \neq 1/2,\\
    \overline{x}(t)\ln\left(N/\overline{x}(t)\right), &\alpha=1/2,
    \end{cases}
\end{align}

\noindent assuming deterministic initial conditions ($\sigma(0) = 0$). Note that using other approximations such as the linear-noise approximation to find $\sigma(t)$ leads to intractable analytics \cite{van1992stochastic}. As shown in Fig.~\ref{fig:5}(a), this approximation does well for a range of parameters, as long as the system remains unimodal (bimodal distributions cannot be approximated by a single Gaussian). As $\alpha$ decreases below $1/2$, and the number distribution becomes more bimodal, the approximation gets gradually worse on the level of distributions. In our case, the breakdown of the WKB theory is an indicator of transiently bimodal behavior.

We expect bimodal behavior when the spread of molecule numbers is large compared to its mean. For discrete systems, this notion is captured by the Fano factor, $\mathrm{FF}(t) = \sigma(t)^2/\overline{x}(t)$ \cite{banerjee2025fano}. The Fano factor is 1 for a Poisson process \cite{van1992stochastic}, $>1$ for a ``bursty process'' \cite{taniguchi2010quantifying}, and $<1$ for more deterministic processes \cite{mcclure1980rate,choubey2015deciphering,weidemann2023minimal}. Given the observed transition point at $\alpha^\star\approx 0.6$ in Fig.~\ref{fig:4}(a) and (c), one might expect to observe discontinuous behaviors in the Fano factor about this point. For $\overline{x}(t) \ll N$, we obtain
\begin{align}\label{eq:FFnum}
    \mathrm{FF}(t)\simeq 
    \begin{cases}
        \left( N/\overline{x}(t) \right)^{1-2\alpha}, &\alpha < 1/2,\\
       \ln(N/\overline{x}(t)), &\alpha = 1/2,\\
        1, &\alpha>1/2.
    \end{cases}
\end{align}


For $\alpha>1/2$, the Fano factor is approximately constant and independent of the initial condition as long as $\overline{x}(t) \ll N$, whereas for $\alpha\leq1/2$, it diverges as the system approaches the absorbing state, $\overline{x}(t)\to 0$. This suggests that the distribution becomes more spread out in this regime, as expected from our observation of bimodality. From a dimensionality perspective, Eq.~\eqref{eq:FFnum} states that for $\alpha\leq1/2$ the initial condition at $N$ is still an important determinant in the molecule number fluctuations: there is a memory of the initial condition even at late stages of the transient's evolution. On the other hand, for $\alpha>1/2$ the Fano factor retains no memory of the initial condition. Therefore, using the WKB approximation, the emergence of transient bimodality can be seen as a second-order phase transition in $\mathrm{FF}(t)$. We note that our prediction of the transition point at \smash{$\alpha_{\mathrm{theory}}^\star=0.5$} slightly underestimates the numerical value of \smash{$\alpha^\star\approx 0.6$} (see Fig~\ref{fig:4}(a)).

With the Fano factor from Eq.~\eqref{eq:FFnum}, we can start connecting transient bimodality to the dynamical behavior of the system. To do so, observe that the acceleration of the system,
\begin{align}
    a(t) := \frac{\mathrm{d}^2\overline{x}(t)}{\mathrm{d}t^2} = \alpha \overline{x}(t)^{2\alpha-1},
\end{align}
has similar behavior to the Fano factor for $\alpha<1/2$:
\begin{align}
    \mathrm{FF}(t)\sim a(\overline{x}(t))/a(N).
\end{align}
Specifically in the regime of $\alpha<0$, the system accelerates as it approaches the absorbing state at $0$. This underlines our narrative for noise-driven transient bimodality: a trajectory that reaches the last non-absorbing stages (the transient regime) earlier than the others will quickly move into the final state, leading to characteristically bimodal distribution (see Fig.~\ref{fig:3}). In other words, outliers from the transiently evolving mode race away to the absorbing state—creating the second mode. Comparing this with Eq.~\eqref{eq:crit}, for $\alpha<0$, the waiting time in the transient regime is small.

Our results also indicate that for $0<\alpha<1/2$ noise-driven transient bimodality is present but is not accompanied by acceleration into the absorbing state. In this regime, where transient bimodality has not previously been reported, the acceleration decreases as 0 is approached, but not fast enough to stop the accumulation of a second mode at 0.

In Fig.~\ref{fig:5}(b), we show how the mean and Fano factor of the system evolve for different values of $\alpha$. For $\alpha=-1$ and $\alpha=0$, the transiently bimodal regime is associated with a pronounced peak in $\mathrm{FF}(t)$. More generally, for $\alpha\leq 0.6$, transient bimodality becomes increasingly more prevalent as $\alpha$ decreases, with $\mathrm{FF}(t)$ rising more rapidly throughout the transiently bimodal regime (blue shaded area). Note that for $\alpha<1/2$, the mean-field approximation predicts that $\mathrm{FF}(t)$ increases monotonically with time $t$; this corresponds to the small molecule number limit wherein the mean-field approximation breaks down (see Fig.~\ref{fig:5}(a)). When $\alpha \approx 0.6$, the transiently bimodal regime becomes essentially a single point in time, i.e., the corresponding peak in $\mathrm{FF}(t)$ is of order $\mathcal{O}(1)$. Finally, at $\alpha=1$, the peak vanishes, and $\mathrm{FF}(t)$ tends to 1. This confirms the expected phenomenology observed in Eq.~\eqref{eq:FFnum}, and validates the divergence of $\mathrm{FF}(t)$ as the final state is approached as a key signature of transient bimodality.

\subsection{First-Passage Times Show Transient Bimodality Becomes Less Prominent for Large Systems}
\label{sec:fpt}

\noindent In view of Eq.~\eqref{eq:crit}, another approach to investigate the presence of transient bimodality in our system uses first-passage times. Here we measure the time it takes the system to go from the initial state $N \gg 1$ to the final state $0$. Heuristically, we define $1$ as the transient state, although our results do not strongly depend on the exact choice. Thus let $\tau_1$ be the time required for the system to reach $1$. As shown in \ref{sec:app_fpt}, this has variance given by
\begin{align}
    \mathrm{Var}(\tau_1) &\sim \begin{cases}
        N^{1-2\alpha}, & \alpha < 1/2, \\
        \log N, & \alpha = 1/2, \\
        1, & \alpha > 1/2,
    \end{cases}
\end{align} 

\noindent for large $N$, up to constants. For any $\alpha$, the system spends an average time of $\overline{\tau_2} = 1$ in the state $1$. \textit{Our criterion for bimodality from Eq.~\eqref{eq:crit} is thus satisfied for large $N$ precisely when $\alpha \leq 1/2$. }

\begin{figure*}[ht]
    \includegraphics[width=\textwidth]{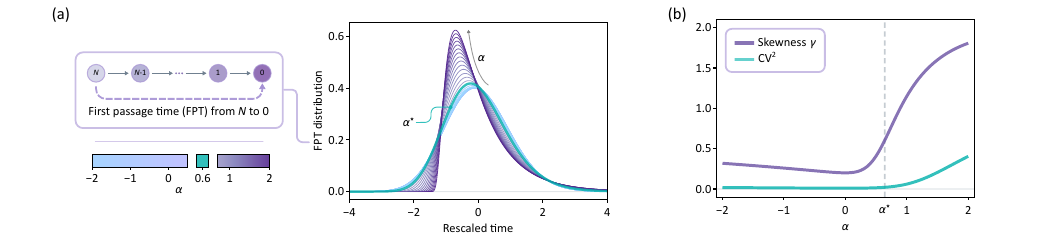}
    \caption{\textbf{First-passage time distributions and statistics across transiently monomodal and transiently bimodal regimes for $N=100$.} (a) Plot of the first-passage time distribution to reach the absorbing state across $\alpha\in [-2,2]$, where the time $t$ is z-score transformed as \smash{$\tilde t = (t-\la t\ra)/\sqrt{\mathrm{Var}(t)}$}. Distributions in purple correspond to transiently monomodal number distributions ($\alpha>0.6$), the turquoise line marks the phase transition point near $\alpha^\star = 0.6$, and the blue lines correspond to transiently bimodal number distributions ($\alpha<0.6$). (b) Plot of the skewness, $\gamma$, and the coefficient of variation squared, $\text{CV}^2$, of the first-passage time distribution. The dashed gray line indicates the phase transition point near $\alpha^\star$, beyond which the skewness and $\text{CV}^2$ increase substantially.}
    \label{fig:FPTs}
\end{figure*}

A more detailed analysis of the first-passage time distribution reveals several key features. First, in the regime $\alpha<1/2$ (in which transient bimodality occurs), the noise in the first-passage time distribution scales as that of a Poisson distribution, becoming small for large $N$. In this regime, the first-passage time distribution becomes approximately Gaussian (see blue curves in Fig.~\ref{fig:FPTs}(a)). For the system in Eq.~\eqref{eq:chain}, this indicates that as $N$ increases, the fraction of time over which transient bimodality occurs becomes increasingly small. We observe this behavior in Fig.~\ref{fig:4}(e), even though the maximal bimodality coefficient during the transient increases with $N$ (Fig.~\ref{fig:4}(d)). 
When $\alpha>1$, the variance in first-passage time distributions remains bounded as $N \rightarrow \infty$. In this regime, the first-passage time distribution tends towards a shifted exponential distribution (Fig.~\ref{fig:FPTs}(a)). This is expected, as for large $\alpha$ the system spends an increasingly large fraction of time in the state $1$. 

Our findings point towards a \textit{critical numerosity}-like phenomenon for $\alpha < 1/2$ \cite{calvert2023critical,calvert2025many,anderson1972more}. A critical numerosity, defined as in \cite{calvert2025many}, is a shift in a collective's \textit{qualitative} behavior at a critical value of system size $N$; e.g., a shift in a distribution's number of modes at a finite value of $N$. For us, the sensitivity in $N$ (representing the initial state of the Markov chain) in the regime of $\alpha\leq1/2$, seen in Fig.~\ref{fig:4}(d) and (e), is reminiscent of a critical numerosity (outside of the steady state regime). Simply by increasing or decreasing the system size, with all else held constant, the system can transition between qualitatively different states---those with transiently bimodal behavior and those without. As a clear example of this, one can compare \cite[Fig.~3(c)]{calvert2025many} to our phase diagrams in Figs.~\ref{fig:4}(a) and (c), both of which emphasize the transition between systems with monomodality versus bimodality.



Further signatures of transient bimodality are found in the higher-order moments of the first-passage time distribution. In Fig.~\ref{fig:FPTs}, we observe that transient bimodality is highly related to distribution shape, with less-skewed first-passage time distributions ($\alpha\leq 1/2$) admitting transient bimodality, while distributions with a heavy skew ($\alpha>1/2$) do not admit transient bimodality. This is related to an observation made by Hathcock and Strogatz in \cite[Fig.~2]{hathcock2022asymptotic}, using a birth-death extinction model, wherein a discontinuity is observed in the skewness of first-passage time distributions at $\alpha=1/2$. In Fig.~\ref{fig:FPTs}(b), we replicate the plot of Hathcock and Strogatz's first-passage time skewness \cite[Fig.~2]{hathcock2022asymptotic} using our death process, and show a similar inflection of the skewness about $\alpha=1/2$. In Section \ref{sec:app_fpt}, we show that the skewness can be approximated by
\begin{align}
    \gamma(
N) \sim 
    \begin{cases}
        N^{-1/2}, &\alpha < 1/3,\\
        N^{-1/2} \ln(N),&\alpha=1/3,\\
        N^{3\alpha - 3/2}, &1/3<\alpha<1/2,\\
        (\ln N)^{-3/2},&\alpha = 1/2,\\
        1, &\alpha>1/2.
    \end{cases}
\end{align}
Importantly, for $\alpha\leq 1/2$ the skewness will tend to zero for large $N$. On the other hand, for $\alpha >1/2$ the skewness will approach a nonzero value in the same limit (see the purple line in Fig.~\ref{fig:FPTs}(b)). Therefore, the regime of transient bimodality is characterized by first-passage time distributions with negligible skewness, in agreement with the results observed for the scaling of the coefficient of variation squared of the first-passage time distribution (see the turquoise line in Fig.~\ref{fig:FPTs}(b)).

Using an illustrative example of a Markov death process with power-law transition rates, we have shown some of the core features of transient bimodality from the perspective of the statistics of the number distribution and first-passage time distribution. Our theory predicts a phase transition into transient bimodality below $\alpha^\star_\mathrm{theory}=1/2$, which is observed at $\alpha^\star\approx 0.6$ for simulations on a chain of length $N=100$. In the next section, we categorize the core features of transient bimodality exposed by our findings.


\subsection{Conditions for Noise-Driven Transient Bimodality}
\noindent Transiently bimodal behavior can occur for a wide variety of systems that do not exhibit deterministic bistability or steady-state bimodality. Our time-dependent analysis of a Markov death process has revealed five key features that are defining of noise-driven transient bimodality:

\begin{enumerate}
    \item Larger systems have the potential to express greater transient bimodality, albeit over a smaller fraction of the time evolution towards the final state (see Fig.~\ref{fig:4}(e)).
    \item Processes that start slow and end fast ($\alpha<0$) express greater transient bimodality (see Fig.~\ref{fig:4}), although processes that start fast and end slow can still admit transient bimodality in the range $0<\alpha\lesssim 0.6$ (theoretically predicted for $0<\alpha<0.5$).
    \item A process with a strong rate-limiting step towards the end of its time evolution will generally not admit transient bimodality (see Fig.~\ref{fig:FPTs}(a)).
    \item A significant indicator of transient bimodality lies in the Fano factor of the number distribution. If the Fano factor increases substantially and mean-field theories break down as the transient evolves, this indicates a transition into a transiently bimodal regime (see Fig.~\ref{fig:5}).
    \item On the level of first-passage times, the conditions for transient bimodality require that the average time to leave the transient state should be less than the standard deviation of the time to reach the transient state (see Eq.~\eqref{eq:crit}).
\end{enumerate}

We pause to make several comments. First, points 1 and 3 above explain why transient bimodality is empirically rarer than transient monomodality in cell biology --- molecule numbers can be of the order of hundreds or thousands, taking place in processes with rate limiting steps (see the wide range of $k_\mathrm{cat}$ values in \cite{davidi2018bird}). Second, our articulation that processes that can be transiently bimodal under fast-to-slow dynamics in point 2 has not previously been stated. Previous literature generally confines the occurrence of transient bimodality to processes that start slow and end fast---many authors have studied deterministic potentials that encode this property \cite{Baras1983,frankowicz1983transient,frankowicz1984stochastic}. Third, similar to the work of \cite{frankowicz1983transient}, for processes with transiently bimodal behavior we see: (i) an initial time period over which transient bimodality is not present; (ii) a marked increase in the skewness of the number distribution; and (iii) a mean behavior that aligns well with that predicted by the deterministic rate equations. 









\section*{Discussion}\label{sec:discussion}

\noindent In this article we have explored the origin and nature of transient bimodality. We derived a general criterion for transient bimodality in Eq.~\eqref{eq:crit}, and illustrated its accuracy on a minimal model of transient bimodality and a general death chain. In contrast to previous hypotheses (e.g., \cite{Baras1983,nicolis1989exploring,van1992stochastic,frankowicz1984stochastic}), our analysis shows that transient bimodality is not generally explained by bistability in deterministic systems, and does not require environmental noise: it is an intrinsically stochastic phenomenon. This can be shown at the level of either the first-passage time or the number distribution. This highlights the role of intrinsic noise in shaping dynamical behavior, even in systems that appear largely deterministic. 

Transient bimodality is ubiquitous in cell biology \cite{palani2012transient,maithreye2008delay,levien2025slow,blake2003noise,blake2006phenotypic,venturelli2015population,stockwell2017living}, laser optics \cite{lange1985study,broggi1984transient}, the collective memory of cockroaches \cite{calvo2023emergence}, and electrical circuits \cite{Iwaniszewski1994}. A specific field with renewed interest in transient behaviors is theoretical ecology, wherein longitudinal data make it clear that many complex ecological behaviors have transient origins \cite{abbott2021transients,vidiella2018exploiting,hastings2018transient,hastings2021effects,abbott2021mapping,vidiella2021habitat,koch2024biological}. Key advances have been made in the understanding of ecological transient behaviors, generally under deterministic frameworks---occasionally perturbed by noise \cite{vidiella2018exploiting,hastings2018transient,hastings2021effects,abbott2021mapping}. Ghosts---memories of prior steady states, that linger around before new steady states are attained---are an important contemporary phenomenon, and their presence complicates the understanding of when environmental tipping points are crossed \cite{vidiella2018exploiting,hastings2018transient}. Although distinct, transient bimodality shares similar characteristics to ghosts, because the trailing mode is a memory of the initial condition. Similar to some of the literature in cell biology \cite{levien2025slow}, some ecological studies have shown that there are complex interactions between deterministic processes and environmental noise in ecological transients \cite{abbott2021mapping,hastings2021effects}. While the relevance of our work to ecology is presently unknown, we suspect that transient bimodality could be as relevant to ecological mechanisms as it is to molecular mechanisms inside the cell.



Our study suffers from two main limitations. First, we currently lack a taxonomy of stochastic dynamical systems as with deterministic systems and bifurcation theory \cite{strogatz2024nonlinear} While previous work has emphasized a taxonomy of transiently bimodal behavior (see \cite{holehouse2026distinct}), understanding how these different mechanisms change a system's behavior remains subject to exploration. Furthermore, experimentally probing transient bimodality in a biological context can be difficult due to the need to gather time-series data at the single-cell level, or across populations in ecology. While our criterion provides a simple way to predict when transient bimodality can emerge, we expect that ongoing advances in data collection across these domains will help us develop a deeper understanding of this phenomenon \cite{fernandes2022synthetic,chen2022live}, building upon the many examples of empirical transient bimodality highlighted in this study.

Biology is an out-of-equilibrium endeavor, and the \textit{processes of development and growth are, by definition, transient}. Taking seriously the prospect of phenomena arising from time-dependent dynamics is therefore a vital aspect of theoretical biological research. For example, in contrast to steady state mechanisms, a benefit of time-dependent phenomena driven by intrinsic noise is that they can occur simultaneously as environments change \cite{holehouse2026distinct}. This indicates that steady state mechanisms of generating phenotypic heterogeneity---known to be hijacked by cancerous mechanisms \cite{barakat2010learning,mortus2014developmental}---may not be required for certain modes of cellular differentiation under transient mechanisms. In fact, transient bimodality has long been hypothesized to serve as a mechanism for cellular bet-hedging \cite{blake2006phenotypic}.  Our study provides initial steps in motivating a taxonomy of time-dependent stochastic behaviors, that may be useful in explaining developmental biological function.

\section*{Data and Code Availability}

\noindent This paper does not generate or directly analyse empirical data. The code for this paper is available at \href{https://github.com/augustinas1/transient-bimodality}{\textcolor{blue}{https://github.com/augustinas1/transient-bimodality}}. 


\section*{Acknowledgments}

\noindent J.H.~would like to acknowledge the support of the National Science Foundation Grant Award Number EF--2133863. K.\"O and A.S acknowledge funding through an Australian Research Council Laureate Fellowship (FL220100005). J.H.~would like to thank João Pedro Teuber Carvalho for bringing to light the experimental work conducted in \cite{levien2025slow} in addition to further discussions on the empirically observed transient bimodality. J.H., A.S., and K.\"O.~would like to thank the Santa Fe Institute for hosting a working group on \textit{``The Origins of Transient Bimodality''.}

\section*{Author Contributions}
\noindent Conceptualization: J.H., with initial inputs from K.\"O.~and A.S., and later inputs from A.K. Mathematical analysis: J.H., and K.\"O. Funding acquisition: J.H. Simulations and computational analyses: J.H., A.S., and K.\"O. Figures: J.H, A.S. and K.\"O. Writing--original draft: J.H. Writing--review and editing: all authors.

\bibliographystyle{naturemag.bst}
\bibliography{main}

\pagebreak
\clearpage
\widetext
\begin{center}
\textbf{\large Supplementary Information}
\end{center}
\setcounter{equation}{0}
\setcounter{figure}{0}
\setcounter{section}{0}
\setcounter{table}{0}
\setcounter{footnote}{0}
\setcounter{page}{1}
\makeatletter
\renewcommand{\thesection}{S\arabic{section}}
\renewcommand{\theequation}{S\arabic{equation}}
\renewcommand{\thefigure}{S\arabic{figure}}


\section{Snapshot distribution}\label{apdx:cme}

\noindent Here we derive the solution of the master equation reported in Eq.~\eqref{eq:me_sol}. While it is possible to derive the solution from first principles (see \cite{smith2015general}), a more intuitive solution utilizes the first-passage time distribution. The first-passage time distribution in Eq.~\eqref{eq:chain}, to start at $N$ and reach state $n$ at a time $t$, is
\begin{align}
    f_n(t)= \sum_{i={n+1}}^{N} \frac{k_i e^{-k_i t}}{\prod_{\substack{j=n + 1\\ j\neq i}}^{N}(1-\frac{k_{i}}{k_{j}})},
\end{align}
which is simply the hypoexponential distribution with parameters $k_i = i^\alpha$. The first-passage time distribution is related to the probability flux into state $n$ via 
\begin{align}
    f_n(t) = k_{n+1}P(n+1,t).
\end{align}

\noindent Therefore, one finds $P(n,t)$ as
\begin{align}
    P(n,t) = \frac{1}{k_n}\sum_{i={n}}^{N} \frac{k_i e^{-k_i t}}{\prod_{\substack{j=n\\ j\neq i}}^{N}(1-\frac{k_{i}}{k_{j}})},
\end{align}
as reported in the main text.

\section{Mean-field approximation}\label{sec:mft}
\noindent In this section, we derive the mean-field theory for the number distribution in Eq.~\eqref{eq:pxt}. Aspects of this solution were originally presented in \cite{Baras1983}, based on analytic techniques formulated in \cite{kubo1973fluctuation}. We provide a slightly more detailed exposition of this derivation below. Our approximation assumes that the system size $N$ is large. We also assume that $P(n,t)$ is approximately Gaussian, centered at the deterministic solution \eqref{eq:mfmean}. 

We use the WKB method for $P(n,t)$ and write
\begin{align}\label{eq:wkb-p}
    P(n,t) = \exp(N \Phi(z,t)),
\end{align}

\noindent where $z = n/N$ can be treated as a continuous variable. Substituting Eq.~\eqref{eq:wkb-p} into Eq.~\eqref{eq:cme} results in a partial differential equation (PDE) for $\Phi$ in $z$ and $t$,
\begin{align}\label{eq:wkb-app}
    \partial_t\Phi \approx \overline{\mu}(z)\left(\exp(\partial_z \Phi) - 1)\right),
\end{align}
where we have defined the rescaled propensity function $\overline{\mu}(z) = k_z/N$. We now expand $z$ around its deterministic value as $z = \overline{z}+\zeta$, where $\overline{z} = \overline{x}/N$ is computed in Sec.~\ref{sec:det}. 
We then assume that $\Phi$ is analytic as a function of $\zeta$:
\begin{align}\label{eq:phi-exp}
    \Phi(\zeta, t) = a_0(t) + a_1(t) \zeta +a_2(t) \zeta^2 + a_3(t) \zeta^3+\cdots.
\end{align}
If we assume that the deterministic trajectory where $\zeta = 0$ is the most likely trajectory, $\partial_\zeta \Phi(0,t) = 0$ implies that $a_1(t) = 0$. Plugging \eqref{eq:phi-exp} into Eq.~\eqref{eq:wkb-app} and expanding $\overline{\mu}$ in a Taylor series around $\overline{z}$ yields
\begin{align}\label{eq:lhs}
    \dot a_0 + (2a_2 \dot\zeta)\zeta +(\dot a_2+3a_3\dot\zeta)\zeta^2+\cdots &= (2\overline\mu(\overline z)a_2)\zeta + [ \overline{\mu} (3a_3+2a_2^2)+2a_2\overline{\mu}'(\overline{z})]\zeta^2+\cdots,
\end{align}
where the dot indicates a time derivative.

We can now proceed by matching orders of $\zeta$ on each side of Eq.~\eqref{eq:wkb-app}. To order $\zeta^0 = 1$, we find $\dot a_0 = 0$, implying that $a_0$ is constant in time. This is not surprising, since $a_0$ is determined by the normalization of $P(n,t)$. To order $\zeta^1 = \zeta$, we find
\begin{align}
    2a_2\dot\zeta = 2a_2\overline\mu(\overline z)\; \Rightarrow \; \dot\zeta = \overline\mu(\overline z),
\end{align}
which agrees with the deterministic rate equation (van Kampen calls this the ``emergence of a macroscopic law'', \cite[Chap.~X]{van1992stochastic}). 
To order $\zeta^2$, we find
\begin{align}\label{eq:a2_pde}
    \dot a_2 = 2a_2^2\overline\mu(\overline z) + 2a_2\overline\mu'(\overline z).
\end{align}
This differential equation corresponds to a hidden Riccati-type ordinary differential equation, parameterized by $\overline z$. To solve this, we first convert the partial time derivative directly 
\begin{align}
    \partial_t = (\partial_t \overline{z}) \partial_{\overline{z} } = -\overline{\mu}(\overline{z})\partial_{\overline{z}},
\end{align}
which converts Eq.~\eqref{eq:a2_pde} into
\begin{align}\label{eq:a2_ode}
    -\overline{\mu}(\overline{z})\frac{\partial a_2}{\partial \overline z} = 2a_2^2+2a_2\frac{\partial \overline{\mu}(\overline{z})}{\partial \overline{z}}.
\end{align}
Now, let $y(\overline{z}) = a_2(\overline{z})\overline{\mu}(\overline{z})^2$. This dependent variable transformation converts Eq.~\eqref{eq:a2_ode} into
\begin{align}
    \frac{\partial y}{\partial \overline{z}} = -\frac{2y^2}{\overline{\mu}(\overline{z})^2}.
\end{align}
Noticing that $\partial(1/y)/\partial \overline z = -1/y^2 \partial y/\partial \overline{z}$, one can show that 
\begin{align}
    \frac{1}{y} = \int_{1}^{\overline z}\frac{2}{\overline \mu(s)^2}\mathrm{d}s.
\end{align}
Undoing the transformation from $a_2\to y$ gives us our solution for $a_2(\overline z)$,
\begin{align}\label{eq:a2sol}
    a_2(\overline z)^{-1} = 2\overline{\mu}(\overline z)^2 \int_1^{\overline z}\frac{\mathrm{d}s}{\overline \mu(s)^2}.
\end{align}
Finally, making the intensive variables extensive again, e.g., $k_{\overline x} = N\overline{\mu}(\overline x)$ and $\overline{x} = N \overline{z}$, we then find
\begin{align}
    a_2(\overline x)^{-1} = 2 N k_{\overline x}^2\int_{N}^{\overline x}\frac{\mathrm{d}s}{\overline k_s^2}.
\end{align}

Bringing together the results from Eqs.~\eqref{eq:wkb-p}, \eqref{eq:phi-exp} and \eqref{eq:a2sol}, gives us our mean-field Gaussian approximation of $P(n,t)$,
\begin{align}
    P(n,t) &= \frac{1}{\sqrt{2\pi \sigma(t)^2}}\exp\left(-\frac{(x-\overline{x}(t))^2}{2\sigma(t)^2}\right),\\
    \sigma(t)^2 &= -\frac{1}{2}Na_2(\overline x)^{-1}.
\end{align}
Now taking our power-law death rate $k_x = x^{\alpha}$, we find that 
\begin{align}
    \sigma(t)^2 = 
    \begin{cases}
    \frac{1}{1-2\alpha}\left[N \left( \frac{\overline{x}(t)}{N} \right)^{2\alpha} - \overline{x}(t) \right], &\alpha \neq 1/2,\\
    \overline{x}(t)\ln\left( N/\overline{x}(t)\right), &\alpha=1/2,
    \end{cases}
\end{align}

 \noindent as reported in the main text.

We test the mean, $\overline x(t)$, and standard deviation, $\sigma(t)$, predicted by this theory in Fig.~\ref{fig:5}(a), for cases corresponding to transient mono- and bi-modality. In general, we find the mean-field theory to be a good approximation of the exact results from Eq.~\eqref{eq:pxt} and \eqref{eq:sigma}, at least until a finite time singularity is approached in $\sigma(t)$ for $\overline x(t)\ll N$. Finite time singularities in the mean-field theory occur for $\alpha<1/2$, and they become increasingly more divergent as $\alpha$ decreases.


\section{First-passage time moments}\label{sec:app_fpt}

\noindent Here we derive the mean, variance, and skewness of the first-passage time distribution for the power-law death process in Section \ref{sec:fpt}. Starting from the state $N$, the system spends a random amount of time $\tau_i$ in the state $i$ for $1 \leq i \leq N$, where $\tau_i \sim \mathrm{Exp}(k_i)$ are independent. If $t_0 = \tau_N + \ldots + \tau_1$ is the time to enter the absorbing state, it follows that
\begin{align}
    \begin{split}
        \la t_0 \ra &= \sum_{i=1}^N \frac 1 {k_i} = \sum_{i=1}^N \frac 1 {i^\alpha},\\\mathrm{Var}(t_0) &= \sum_{i=1}^N \frac 1 {k_i^2}  = \sum_{i=1}^N \frac 1 {i^{2\alpha}}, \\
        \la (t_0 - \la t_0 \ra)^3 \ra &= 2 \sum_{i=1}^N \frac 1 {k_i^3} = 2 \sum_{i=1}^N \frac 1 {i^{3\alpha}}.
    \end{split}
\end{align}

\noindent We approximate these sums by integrals using the midpoint rule:
\begin{align}
    \begin{split}
    \la t_0 \ra &\approx \int_{1/2}^{N+1/2} \frac 1 {s^\alpha} \, ds = \frac 1 {1 - \alpha} \left(\left(N + \frac 1 2\right)^{1 - \alpha} - \left(\frac 1 2\right)^{1 - \alpha}\right), \\ 
    \mathrm{Var}(t_0) &\approx \int_{1/2}^{N+1/2} \frac 1 {s^{2\alpha}} \, ds = \frac 1 {1 - 2\alpha} \left(\left(N + \frac 1 2\right)^{1 - 2\alpha} - \left(\frac 1 2\right)^{1 - 2\alpha}\right), \\
    \la (t_0 - \la t_0 \ra)^3 \ra &\approx 2\int_{1/2}^{N+1/2} \frac 1 {s^{3\alpha}} \, ds = \frac 2 {1 - 3\alpha} \left(\left(N + \frac 1 2\right)^{1 - 3\alpha} - \left(\frac 1 2\right)^{1 - 3\alpha}\right).
    \end{split}
\end{align}

\noindent As a result, we get for the skewness
\begin{align}
    \gamma &\approx \frac {2 (1 - 2\alpha)^{3/2}} {1 - 3\alpha} \frac{\left(N + \frac 1 2\right)^{1 - 3\alpha} - \left(\frac 1 2\right)^{1 - 3\alpha}}{\left(\left(N + \frac 1 2\right)^{1 - 2\alpha} - \left(\frac 1 2\right)^{1 - 2\alpha}\right)^{3/2}} = \frac {2^{3/2} (1 - 2\alpha)^{3/2}} {1 - 3\alpha} \frac{\left(2 N + 1\right)^{1 - 3\alpha} - 1}{\left(\left(2N + 1\right)^{1 - 2\alpha} - 1\right)^{3/2}}.
\end{align}

\end{document}